\documentclass[prd,amsmath,showpacs,preprintnumbers,%
               superscriptaddress,floatfix,twocolumn,letterpaper]{revtex4}

\usepackage{graphicx}

\newcommand{\oa}[1]{\ensuremath{{\cal O}(a^{#1})}}

\newcommand{\eref}[1]{Eq.~(\ref{#1})}

\begin{document}

\preprint{ADP-07-03/T643}

\title{Scaling behavior and positivity violation of the gluon
  propagator in full QCD}

\author{Patrick O.\ Bowman}
\affiliation{Centre of Theoretical Chemistry and Physics, 
  Institute of Fundamental Sciences, Massey University (Auckland), 
  Private Bag 102904, NSMSC, Auckland NZ}
\author{Urs M.\ Heller}
\affiliation{American Physical Society, One Research Road, Box 9000, Ridge, 
  NY 11961-9000, USA}
\author{Derek B.\ Leinweber}
\affiliation{Special Research Centre for the Subatomic Structure of Matter 
  (CSSM), Department of Physics, University of Adelaide 5005, Australia}
\author{Maria B.\ Parappilly}
\affiliation{Special Research Centre for the Subatomic Structure of Matter 
  (CSSM), Department of Physics, University of Adelaide 5005, Australia}
\affiliation{School of Chemistry, Physics and Earth Sciences, Flinders
  University, GPO Box 2100, Adelaide 5001, Australia}
\author{Andr\'e Sternbeck}
\author{Lorenz von Smekal}
\author{Anthony G.\ Williams} 
\affiliation{Special Research Centre for the Subatomic Structure of Matter 
  (CSSM), Department of Physics, University of Adelaide 5005, Australia}
\author{Jianbo Zhang}
\affiliation{Department of Physics, Zhejiang University, Hangzhou, 
  Zhejiang 310027, P.R.\ China}

\date{October 22, 2007}

\begin{abstract}
  The Landau-gauge gluon propagator is studied using the coarse and fine 
  dynamical MILC configurations.  The effects of dynamical quarks are clearly
  visible and lead to a reduction of the nonperturbative infrared enhancement 
  relative to the quenched case.  Lattice spacing effects are studied and found
  to be small. The gluon spectral function is shown to clearly violate
  positivity in both quenched and full QCD. 
\end{abstract}

\maketitle

\section{Introduction}

Quarks and gluons, which carry color charge, are not observed as free 
particles, but only inside the colorless bound states called hadrons.  
This phenomenon is called confinement. The gluon propagator has long 
been studied regarding confinement. It is also an important input for
many phenomenological calculations in hadronic physics 
(see e.g.\ the reviews
\cite{Roberts:1994dr,Roberts:2000aa,Alkofer:2000wg,Maris:2003vk,Holl:2006ni,
  Fischer:2006ub}). 

In the last decade there have been many lattice studies devoted to the
gluon propagator in the Landau gauge. Most of them are done either in
quenched QCD \cite{Mandula:1987rh,Bernard:1994tz,Marenzoni:1995ap,Ma:1999kn,Becirevic:1999uc,Becirevic:1999hj,Nakajima:2000nr,Bonnet:2001uh,Langfeld:2001cz,Leinweber:1998uu,Leinweber:1998im,Bonnet:2000kw,Sternbeck:2005tk,Silva:2005hb} 
or in quenched $SU(2)$ \cite{Bloch:2003sk,Cucchieri:1998fy,Cucchieri:2003di}, 
but there are also more recent reports
\cite{Bowman:2004jm,Sternbeck:2006rd,Ilgenfritz:2006gp,Ilgenfritz:2006he} by us 
and others for the case of full QCD. All lattice studies so far have
indicated that the gluon propagator in Landau
gauge is infrared finite (see
e.g.\ \cite{Bonnet:2001uh,Bowman:2004jm,Sternbeck:2006cg}). 
This, however, disagrees with the corresponding results 
obtained in studies of the Dyson-Schwinger equation (DSE) for the
gluon propagator. There an infrared vanishing gluon propagator is
predicted \cite{vonSmekal:1997is,vonSmekal:1997vx,Lerche:2002ep,Alkofer:2000wg}.
Such an infrared behavior has also been found using stochastic 
quantization \cite{Zwanziger:2001kw} and in studies of exact 
renormalization group equations \cite{Pawlowski:2003hq}.
Recent DSE studies on a finite torus strongly suggest that this discrepancy is 
due to finite volume effects~\cite{Fischer:2007pf}.

In this paper we extend the previous work reported
in~\cite{Bowman:2004jm} to finer lattices studying the scaling
behavior of the gluon propagator in Landau gauge. We
show that the gluon Schwinger function in quenched and full QCD is
negative in a certain interval. This is consistent with both a recent 
lattice study of two flavor QCD \cite{Sternbeck:2006cg} and with
results obtained in DSE studies \cite{Alkofer:2000wg,Alkofer:2003jj}. 

We use a one-loop Symanzik improved gauge
action~\cite{Luscher:1984xn}. It was noted some time ago (e.g.\
Ref.~\cite{Luscher:1984is}) that improved gauge actions can lead to
violations of positivity. These violations are lattice artifacts and
vanish in the continuum limit. In contrast, the positivity violations
of the gluon propagator reported here are independent of
the lattice spacing and are therefore expected to survive the
continuum limit. They are not associated with the choice of action.

\section{Positivity of Euclidean Green's functions}
\label{sec:positivity_of_euclideanGF}

Here, a note on positivity and its violation is in
order. In quantum field theory in Minkowski space, if a certain degree
of freedom is supposed to describe a physical asymptotic state, it
must not have any negative norm contributions in its propagator. That
is, the propagator must not violate positivity. Otherwise the states it 
describes cannot be part of the physical state space; they are,
 so to say, confined from the physical world.

Considering Euclidean Green's functions, positivity translates into
the notion of reflection positivity as one of the famous
Osterwalder-Schrader axioms
\cite{Osterwalder:1973dx,Osterwalder:1974tc} of Euclidean 
quantum field theory (see also \cite{Haag:1992hx,Glimm:1987ng}). For our 
purpose it is instructive, and also sufficient, to consider reflection
positivity in the case of an Euclidean $2$-point function,
$D(x-y)$, whose corresponding propagator in momentum space, $D(q^2)$,
can be written in a spectral representation 
\begin{equation}
D(q^2)= \int\limits_{0}^{\infty} dm^2\, \frac{\rho(m^2)}{q^2+m^2}
\label{eucliprop}
\end{equation}
where $q^2$ denotes the four-momentum squared in Euclidean space and
$m^2$ is the mass squared. The spectral
function, $\rho(m^2)$, is unknown in general, but if $\rho(m^2)\ge0$ for
all $m^2$, \eref{eucliprop} is known 
as the K\"allen-Lehmann representation.

The 1-dimensional Fourier transform of $D(q^2)$ at zero spatial
momentum defines the wall-to-wall correlator \cite{Aubin:2003ih}
\begin{equation}
C(t) = \int\limits_{0}^{\infty}\, dm\, 
               \rho(m^{2})\, e^{-m t}. 
\label{eq:schwinger}
\end{equation}
In lattice spectroscopy the exponential decay of this function is used to 
extract the mass of a particle state. 

Obviously, from \eref{eq:schwinger}, 
if the spectral function is positive, then $C(t)\geq 0$. This needs
not to be the case the other way around. If, however,
$C(t)$ is found to be negative, $\rho(m^2)$ cannot be a
positive spectral function. That is, there is no K\"allen-Lehmann
representation and the corresponding states cannot appear
in the physical particle spectrum: they are confined.

Alternatively, an effective mass
\cite{Mandula:1987rh,Aubin:2003ih}
\begin{displaymath}
  m_{\textrm{eff}}(t) := - \frac{d}{dt}\ln C(t)
\end{displaymath}
could be considered. For positive $\rho$, the slope of
$m_{\textrm{eff}}(t)$ cannot be positive. If, on the contrary,
$m_{\textrm{eff}}(t)$ increases with $t$, $\rho$ cannot be a positive
spectral function.

The gluon \mbox{2-point} function was known from the beginning to
violate positivity. Already in the first numerical study 
\cite{Mandula:1987rh} of the gluon propagator in quenched
lattice QCD in Landau gauge this has been demonstrated by an effective
gluon mass rising with increasing distance $t$ (see also
\cite{Bernard:1992hy,Marenzoni:1993td,Aiso:1997au}). A direct
observation, however, of a non-positive gluon correlator,
$C(t)$, in particular for full QCD, has been unfeasible for a long time.
Nevertheless, in more recent times a non-positive $C(t)$ has been
observed in studies of the corresponding Dyson-Schwinger  
equation~\cite{Alkofer:2003jj}, in quenched $SU(2)$ lattice gauge theory 
in three dimensions~\cite{Cucchieri:2004mf} and hinted at in QCD with dynamical
quarks~\cite{Furui:2004bq}. Explicit evidence in
quenched and full QCD (clover-improved Wilson fermions) has been 
given recently in \cite{Sternbeck:2006cg}. To complete those findings,
in this paper we show that ${C(t)<0}$ for some $t$ in full QCD with
2+1 flavors of \oa{2} Symanzik-improved staggered fermions.

\section{Gluon propagator on the lattice}

In the continuum, the Euclidean gluon propagator in Landau gauge has
the tensor structure
\begin{equation}
D^{ab}_{\mu\nu}(q) = \left ( \delta_{\mu\nu}-\frac{q_{\mu}q_{\nu}}{q^2} \right) 
   \delta^{ab}D(q^2) \, .
\label{eq:Landau-Prop}
\end{equation}
Here $D(q^2)$ is a scalar function which contains the whole
nonperturbative information of $D^{ab}_{\mu\nu}(q)$. At tree-level 
\begin{equation}
D(q^2) = \frac{1}{q^2}.
\end{equation}
Assuming \eref{eq:Landau-Prop} to be well satisfied on the lattice, we  
have calculated $D(q^2)$ using a variety of quenched and dynamical
configurations. These configurations were provided to us by the MILC
collaboration~\cite{Aubin:2004wf} through the Gauge Connection~\cite{nersc}
and were generated with the \oa{2} one-loop Symanzik improved gauge
action~\cite{Luscher:1984xn}. For the dynamical configurations the
``AsqTad'' quark action was used. This is an \oa{2} Symanzik-improved
staggered fermion action with 2+1 flavors implemented using the
``fourth-root trick''.  The lattice spacing was determined from the variant
Sommer parameter, $r_1$~\cite{Aubin:2004wf}.  Following MILC convention we 
refer to the $28^3\times 96$ configurations as being 
``fine'' lattices and the $20^3\times 64$ configurations as the
 ``coarse'' lattices.  See Table~\ref{simultab} for details.

With this lattice gauge action the Landau-gauge gluon propagator at
tree-level is 
\begin{equation}
D^{-1}(p_\mu) = \frac{4}{a^2}\sum_{\mu}\left\{ 
   \sin^2 \left( \frac{p_\mu a}{2} \right)
   + \frac{1}{3}\sin^4 \left( \frac{p_\mu a}{2} \right) \right\},
\label{eq:imp_tree}
\end{equation}
where 
\begin{equation}
p_\mu  = \frac{2 \pi n_\mu}{a L_\mu}, \qquad
n_\mu \in  \left( -\frac{L_\mu}{2}, \frac{L_\mu}{2} \right],
\label{eq:qhat}
\end{equation}
$a$ is the lattice spacing and $L_\mu$ is the length of the lattice in
the $\mu$ direction. To ensure the correct tree-level behavior for the 
lattice gluon propagator, a ``kinematic'' choice of momentum,
\begin{equation}
q_\mu(p_\mu) \equiv \frac{2}{a}\sqrt{ 
   \sin^2 \left( \frac{p_\mu a}{2} \right)
   + \frac{1}{3}\sin^4 \left( \frac{p_\mu a}{2} \right) }\;.
\end{equation}
is employed~\cite{Bonnet:2001uh}.

\begin{table}[t!]
\caption{\label{simultab} Lattice parameters used in this study. The
dynamical configurations each have two degenerate light quarks
(up/down) and a heavier quark (strange).  In the table we show the sea
quark masses both in dimensionless (lattice) units and estimated physical
units.  For details see Ref.~\cite{Aubin:2004wf}.}
\begin{ruledtabular}
\begin{tabular}{cccc@{\quad}ccc}
   & Dimensions      & $\beta$ & $a$ (fm)&$ma$& $m$ (MeV) & \#Config. \\   
\hline
1  & $28^3\times 96$ &   8.40  & 0.086  &\multicolumn{2}{c}{---
  quenched ---}& 150 \\  
2  & $28^3\times 96$ &   7.09  & 0.086  & 0.062, 0.031 & 14, 68 & 108 \\
3  & $28^3\times 96$ &   7.11  & 0.086  & 0.124, 0.031 & 27, 68 & 110 \\*[1ex]
4  & $20^3\times 64$ &   8.00  & 0.120 &\multicolumn{2}{c}{---
  quenched ---}& 192 \\  
5  & $20^3\times 64$ &   6.76  & 0.121 & 0.010, 0.050 & 16, 82 & 203 \\ 
6  & $20^3\times 64$ &   6.79  & 0.120 & 0.020, 0.050 & 33, 82 & 249 \\ 
7  & $20^3\times 64$ &   6.81  & 0.120 & 0.030, 0.050 & 49, 82 & 268 \\
8  & $20^3\times 64$ &   6.83  & 0.119 & 0.040, 0.050 & 66, 83 & 318
\end{tabular}
\end{ruledtabular}
\end{table}

Lattice Monte Carlo estimates for the bare gluon propagator,
$D(q^2)$, have to be renormalized. The renormalized propagator
$D_R(q^2;\mu^2)$ is related to $D(q^2)$ through
\begin{equation}
D(q^2) = Z_3(\mu^2,a) \, D_R(q^2;\mu^2) \, 
\label{eq:renorm-def}
\end{equation}
where $\mu$ is the renormalization point. The renormalization constant  
$Z_3$ depends on the renormalization prescription. We choose the 
momentum space subtraction (MOM) scheme where $Z_3(\mu^2,a)$ is
determined by the tree-level value of the gluon propagator at the
renormalization point, i.e.\
\begin{equation}
D_R(q^2)\bigg|_{q^2=\mu^2} = \frac{1}{\mu^2} \,.
\label{eq:renorm-mom}
\end{equation}
In this study, we will choose to renormalize either at ${\mu=1}$ or at 
${\mu=4}$~GeV.

\section{Results}

\subsection{The effects of dynamical sea-quarks}

\begin{figure}[t]
  \centering
  \includegraphics[height=0.99\hsize,angle=90]{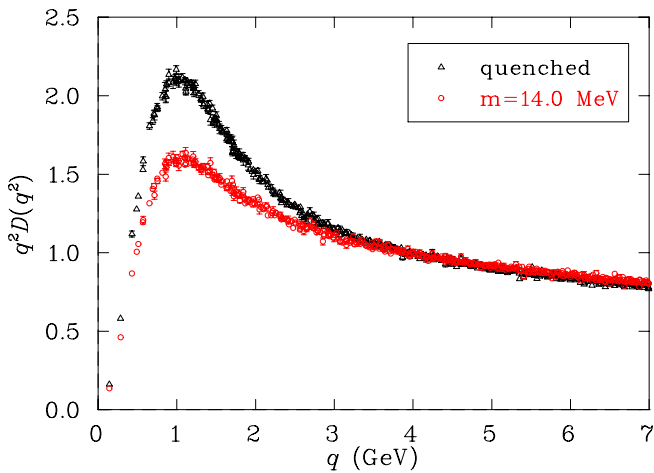}
  \caption{Gluon dressing function in Landau gauge for the fine lattices 
    ($28^3\times 96$).  Full triangles correspond to the quenched calculation, 
    while open circles correspond to 2+1 flavor QCD.  The bare light quark mass
    is $m=14$~MeV for the full QCD result.  As the lattice spacing and volume 
    are the same, the difference between the two results is entirely due to the
    presence of quark loops.  The renormalization point is at $\mu = 4$ GeV.  
    Data has been cylinder cut \protect\cite{Leinweber:1998uu}.}
  \label{gp01}
\end{figure}

To begin with, we discuss the effects of dynamical sea-quarks on the
gluon propagator.  First we compare the dressing function, $q^2D(q^2)$,
in full QCD to that in quenched QCD.  For the fine lattices this is shown
in Fig.~\ref{gp01}. Obviously, 
the infrared hump seen in the quenched case is somewhat suppressed
(about 30\%) due to color screening by the quark--anti-quark pairs.
The basic shape, however, is the same in the quenched and dynamical
cases.  This is consistent with previous results on smaller, coarser 
lattices~\cite{Bowman:2004jm} by some of us, and also
with results obtained independently using clover-improved Wilson fermions
\cite{Sternbeck:2006rd,Ilgenfritz:2006he}.

\begin{figure}[t]
  \centering
  \includegraphics[height=0.99\hsize,angle=90]{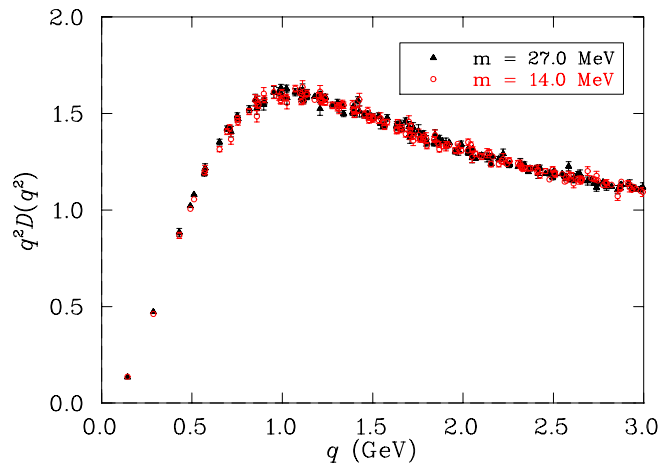}
  \caption{The sea quark mass dependence of the Landau gauge gluon dressing 
    function for the fine lattices.  Filled triangles correspond to bare light 
    quark mass $m=27$~MeV, open circles correspond to the bare light quark
    mass $m=14$~MeV. Data has been cylinder cut
    \protect\cite{Leinweber:1998uu}. No mass dependence is observed
    for this case.} 
  \label{gp02} 
\end{figure}

In Fig.~\ref{gp02} we are looking for the dependence of the dressing function 
on the sea quark mass.  In this case the light quark masses differ by a factor
of two (the strange quark has the same mass in both cases), but we see no 
effect. Note that in Ref.~\cite{Bowman:2004jm} a small quantitative
difference has been observed by studying a slightly greater range of masses
(the heaviest had four times the mass of the
lightest). It would be interesting to study the gluon propagator with
heavier sea quarks so that the transition between quenched and full
QCD could be observed. We leave this for a future paper. 

\begin{figure}[b]
\centering\includegraphics[height=0.99\hsize,angle=90]{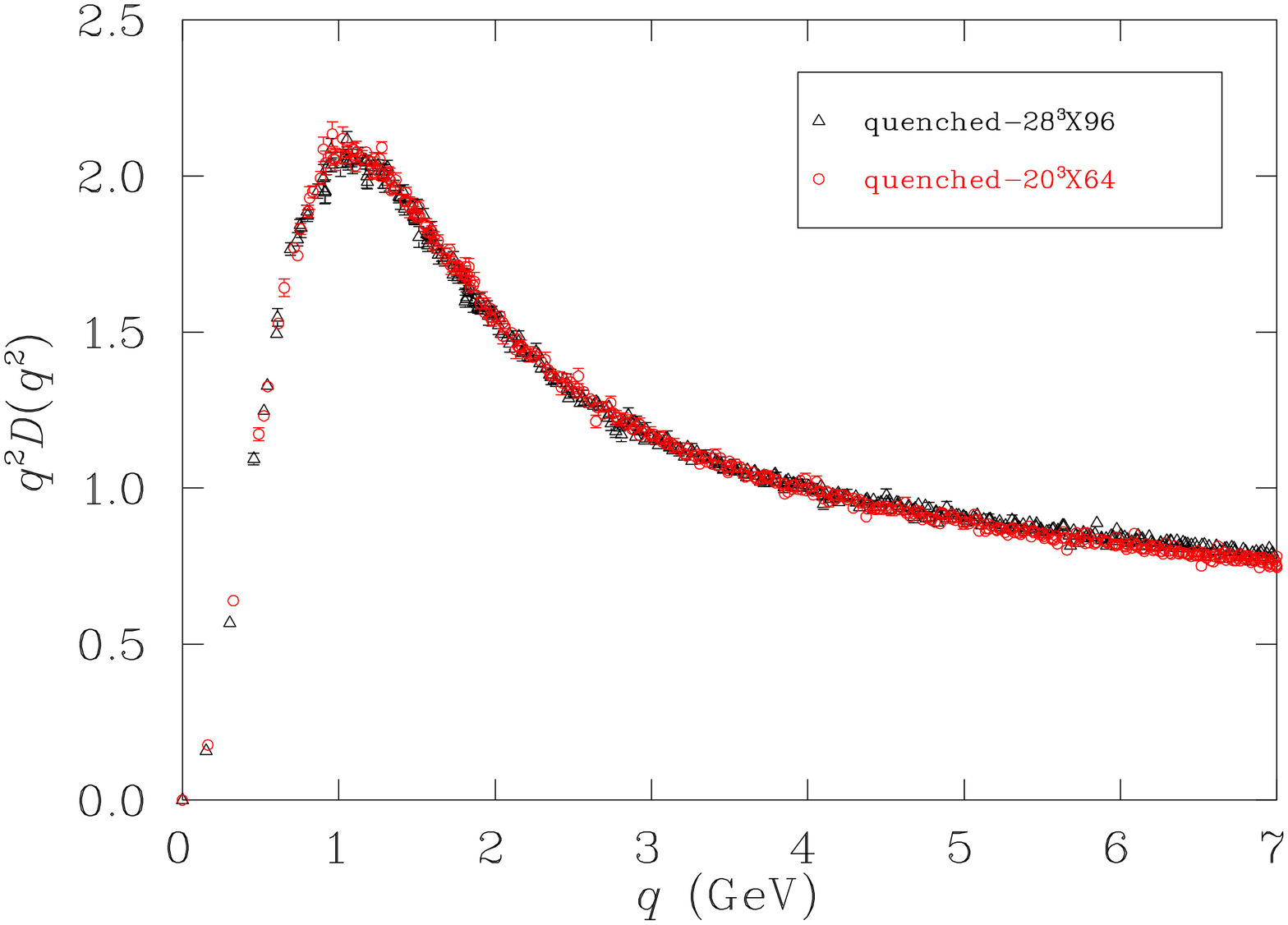}
\caption{The scaling behavior of the quenched gluon dressing function 
renormalized at $\mu = 4$ GeV.  Triangles corresponds to the
gluon propagator on the fine ensemble, open circles to the coarse ensemble.
Good scaling is observed.}
\label{gp03}
\end{figure}

\subsection{The scaling behavior}

The renormalized propagator becomes independent of the lattice spacing as the 
continuum limit is approached. We can see this expected result in the
renormalized quenched dressing function in Fig.~\ref{gp03}. There data
from the coarse and fine lattices are compared, having been 
renormalized at $\mu = 4$ GeV. These two sets of data almost lie on the same 
curve, and hence we conclude that good scaling is found for the quenched 
results. 

\begin{figure}[t]
  \centering
  \includegraphics[height=0.99\hsize,angle=90]{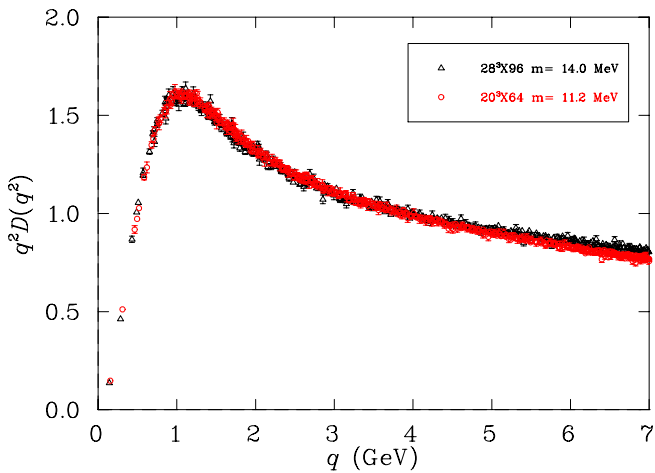}
  \caption{The scaling behavior of the renormalized propagator in full QCD at  
    $\mu = 4$ GeV.  Triangles corresponds to gluon dressing function from the 
    fine ensemble for sea quark mass ${m = 14}$~MeV.  The open circles
    is for sea quark mass ${m = 11}$~MeV which is obtained by extrapolating the
    coarse ensemble data.  Some difference is seen in the large momentum 
    region.}
  \label{gp05}
\end{figure}

\begin{figure}[b]
  \centering
  \includegraphics[height=0.99\hsize,angle=90]{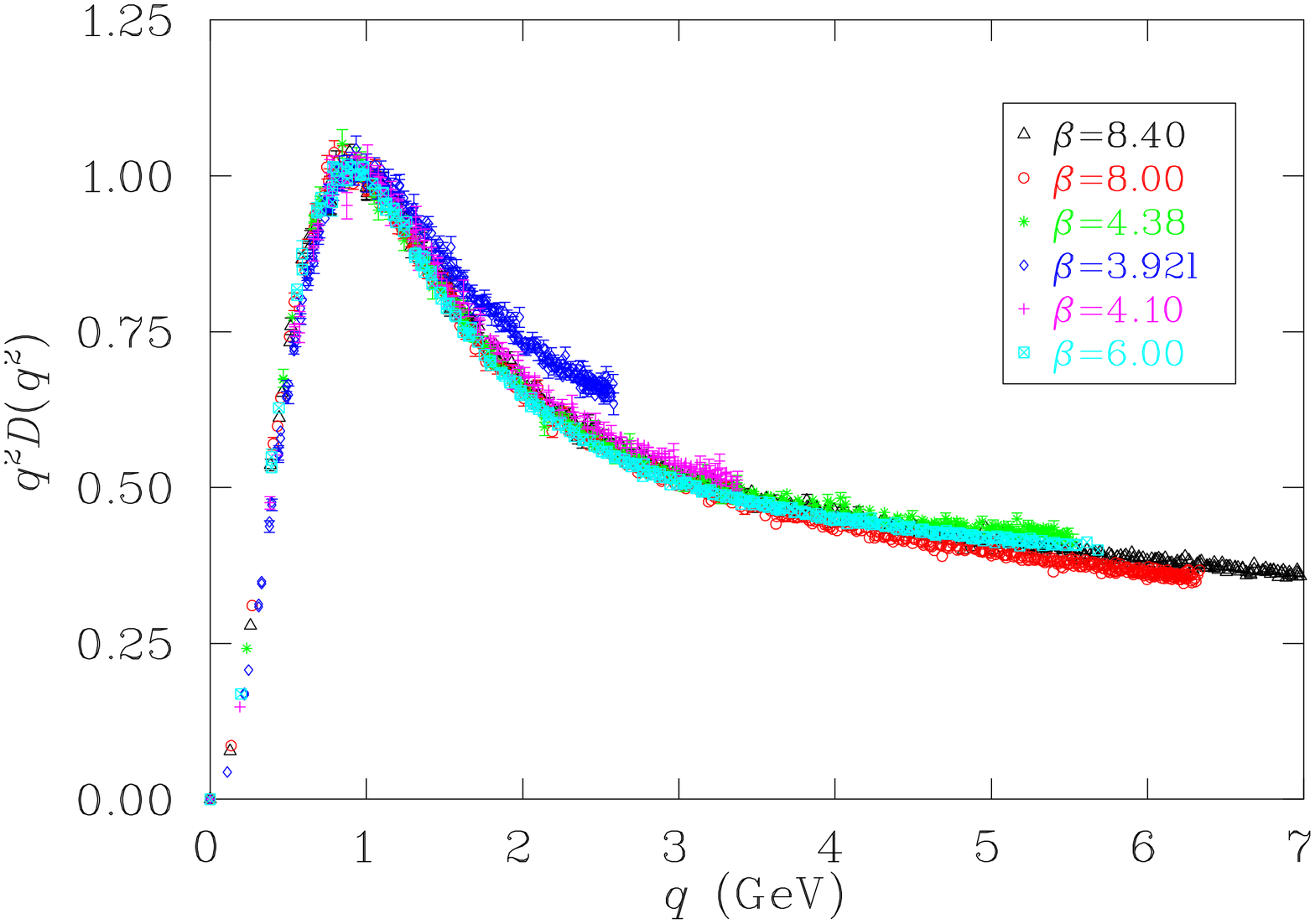}
  \caption{The quenched gluon dressing function renormalized at 1~GeV using
    a variety of lattice spacings and gauge actions.}
  \label{gp05a}
\end{figure}

Turning now to full QCD, in Fig.~\ref{gp05} we compare the gluon dressing 
function from the fine ensemble for the lightest available sea quark mass 
(14~MeV) with that at an approximately equivalent quark mass from the
coarse ensemble. 
Ideally while altering the lattice spacing we should hold the running quark 
mass (rather than bare quark mass) constant.  The quark mass function was 
studied in Ref.~\cite{Parappilly:2005ei} using these gauge configurations and
it was found that a bare quark mass of 14 MeV on the fine ensemble had 
the same running mass as a bare quark of 11 MeV on the coarse ensemble.  We
therefore perform the small extrapolation of the coarse ensemble gluon 
propagator to 11 MeV for this comparison.
Some very small systematic differences at the highest momenta are suggested, but 
this is where the discretization errors are expected to be large.

\begin{table}[t]
\caption{\label{tab2} Quenched configurations used in Ref.~\cite{Bonnet:2001uh}
listed in order of increasing volume.  For this study the lattice spacing has 
been set using the string tension $\sqrt{\sigma} = 440~\text{MeV}$.}
\begin{ruledtabular}
\begin{tabular}{cccc@{\quad}ccc}
   & Dimensions      & $\beta$ & $a$ (fm)& Action & \#Config. \\   
\hline
1  & $16^3\times 32$ &   4.38  & 0.165  & Tree-level Symanzik & 100 \\  
2  & $32^3\times 64$ &   6.00  & 0.0982 & Wilson              &  75 \\
3  & $12^3\times 24$ &   4.10  & 0.270  & Tree-level Symanzik & 100 \\
4  & $10^3\times 20$ &   3.92  & 0.35  & Tree-level Symanzik & 100 \\
5  & $16^3\times 32$ &   3.92  & 0.35  & Tree-level Symanzik & 100 \\  
\end{tabular}
\end{ruledtabular}
\end{table}

In the quenched case we can compare with a wider range of data sets.  
In Fig.~\ref{gp05a} we include results from Ref.~\cite{Bonnet:2001uh}.  These 
were produced using the \oa{2} Symanzik tree-level improved gauge action, 
except for the $\beta=6.0$ configurations which were generated with the 
single-plaquette Wilson gauge action.  We renormalize at 1~GeV so as to 
accommodate the coarsest lattices.  We use the string tension, 
$\sqrt{\sigma} = 440~\text{MeV}$, to set the lattice spacing.
For $\beta=6.0$ the precise measurement of Ref.~\cite{Edwards:1997xf} was used.
A summary of these details is provided in Table.~\ref{tab2}.
The three gauge actions approach the continuum limit slightly differently: for 
the tree-level improved gauge action the 
ultraviolet tail of the gluon propagator drops as the lattice spacing shrinks 
while for the one-loop improved action the tail rises.  They do appear to be
converging to the same result.

\begin{figure}[b]
  \centering
  \includegraphics[height=0.99\hsize,angle=90]{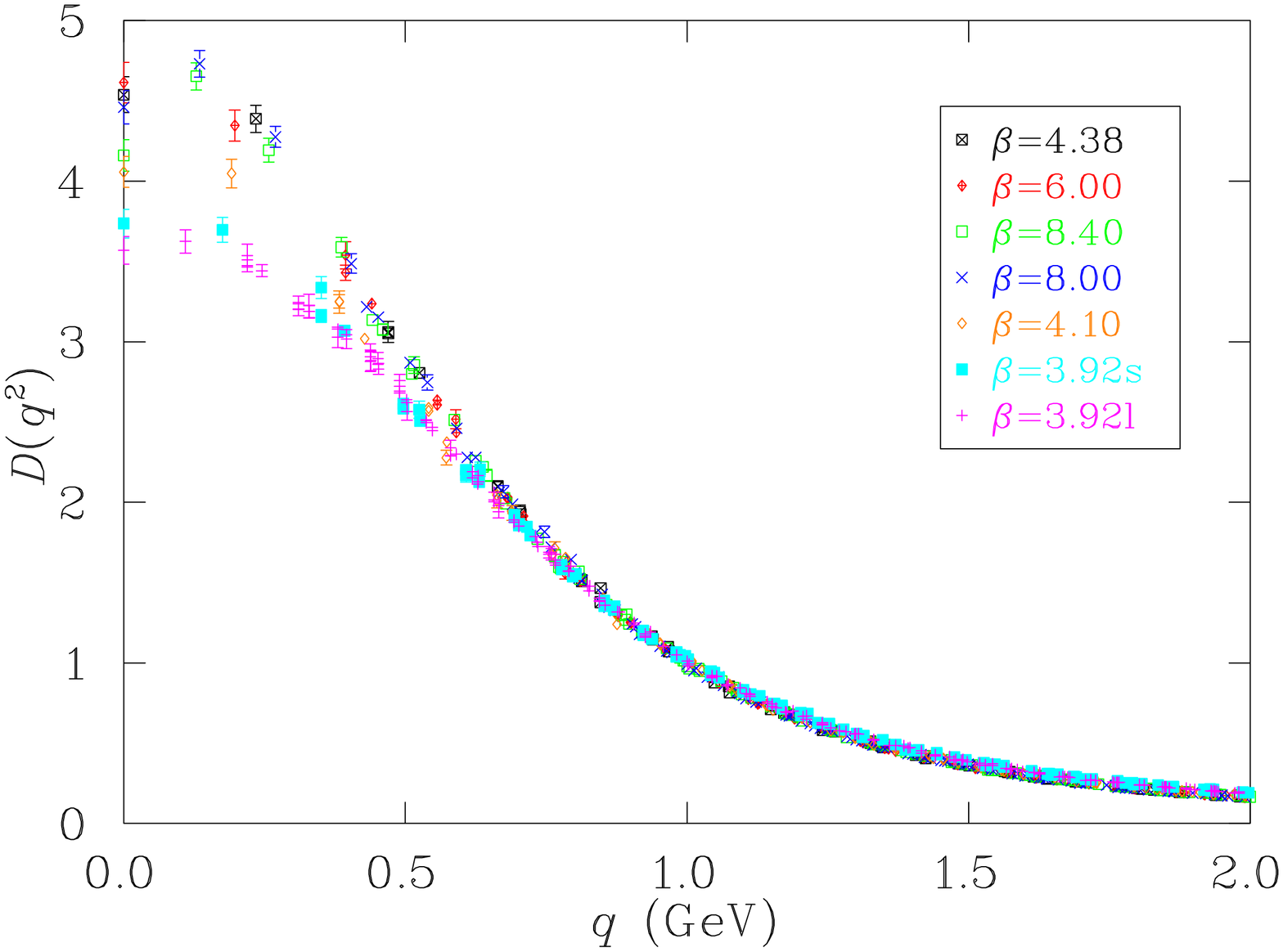}
  \caption{The quenched gluon propagator renormalized at 1~GeV using a variety
    of lattice spacings and three different gauge actions.  The value at
    zero four-momentum drops as the lattice volume increases.}
  \label{gp05b}
\end{figure}

We also revisit the deep infrared region of the gluon propagator itself in
Fig.~\ref{gp05b}.  The $16^3\times 32, \beta=3.92$ lattice has by far the 
largest physical volume ($4.93^3 \times 9.86~\text{fm}^4$) and the smallest
value for $D(q^2=0)$.  The propagator systematically increases as the 
four-volume gets smaller.  Finite volume effects are no longer significant 
for momenta above about 700~MeV.

\begin{figure}[b]      
\centering
\includegraphics[height=0.98\hsize,angle=90]{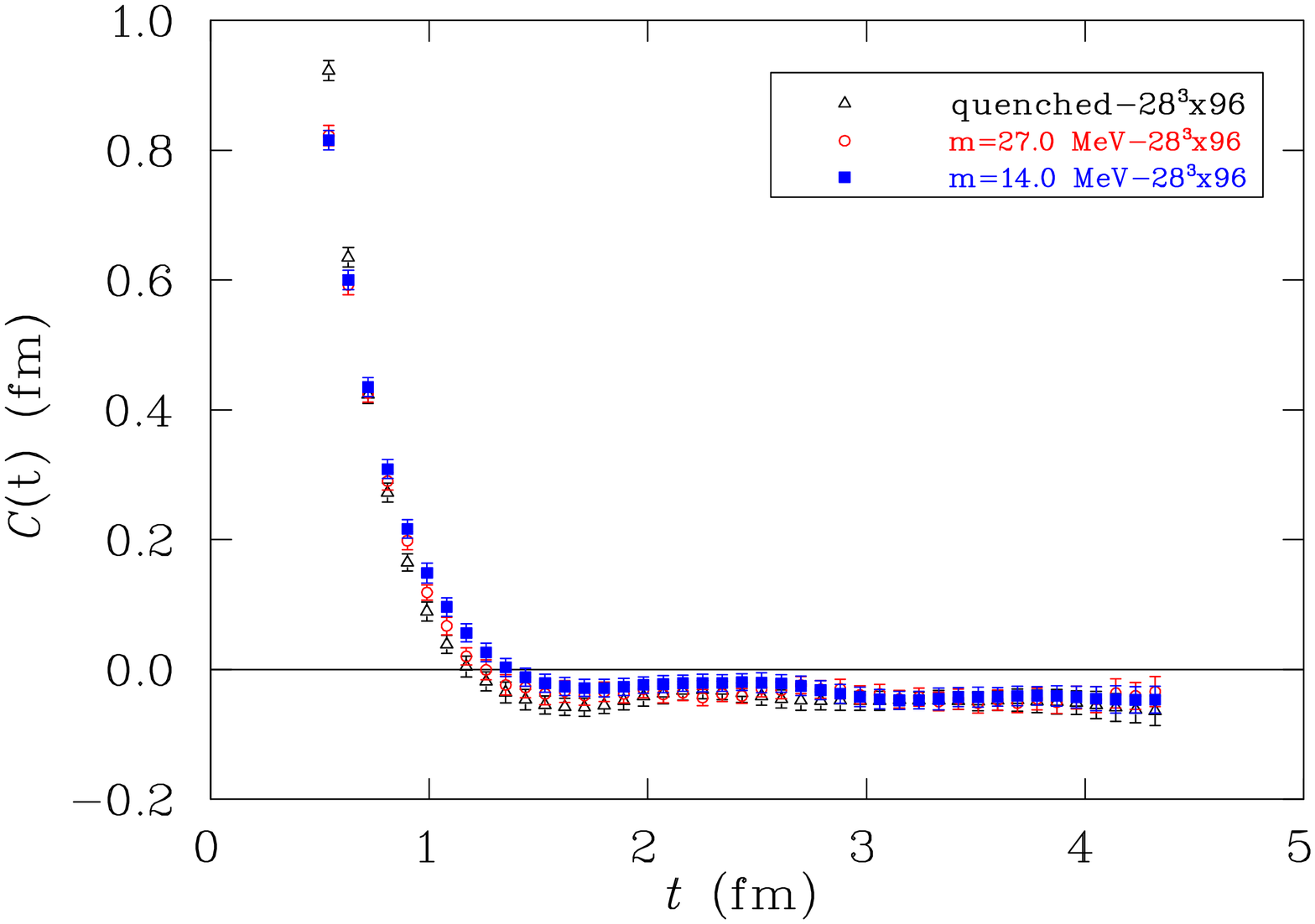}
\caption{The real space propagator $C(t)$ for $28^3\times 96$ lattices plotted 
as a function of dimensionful $t$ for both light sea quark masses and 
the quenched case. Reflection positivity is clearly violated in all
three cases.}
\label{gp06}
\bigskip
\includegraphics[height=0.98\hsize,angle=90]{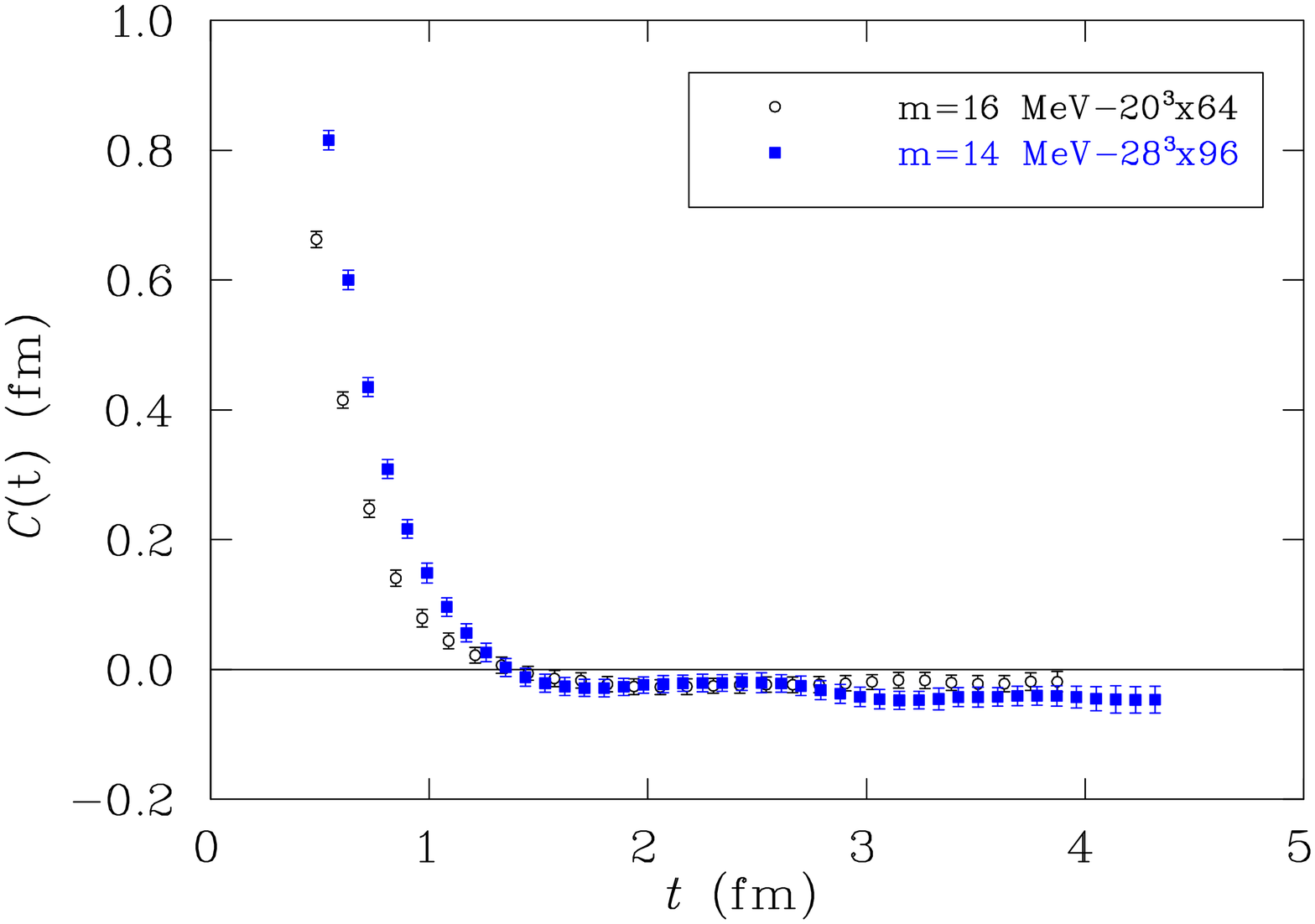}
\caption{The real space propagator $C(t)$ for the coarse and fine
  lattices at approximately same light sea quark masses.}
\label{fig:cmp_Ct_fine2coarse}
\end{figure}

\subsection{Violation of reflection positivity}

Now we turn to the Euclidean time correlator, \eref{eq:schwinger}. On
the lattice this function can be evaluated using the discrete Fourier
transform 
\begin{equation}
  C(t) =  \frac{1}{\sqrt{L_4}} 
  \sum_{n_4=0}^{L_4-1} e^{-2\pi i n_4 t /L_4} \,D(p_4, 0) \;,
\label{eq:ctlattice}
\end{equation}
where $L_4$ is the number of lattice points in time direction,
$p_4$ is the Euclidean time component of the lattice momentum, $n_4$
is an integer and $D(p_4,0)$ is the gluon propagator in momentum space
at zero spatial momentum. 

In Fig.~\ref{gp06}, $C(t)$ is shown for the fine lattices in the
quenched case and for both choices of sea quark. In all cases the
Schwinger function clearly becomes negative signaling
explicit reflection-positivity violation by the gluon propagator.

\begin{figure}[b]
\centering
\includegraphics[height=6.3cm]{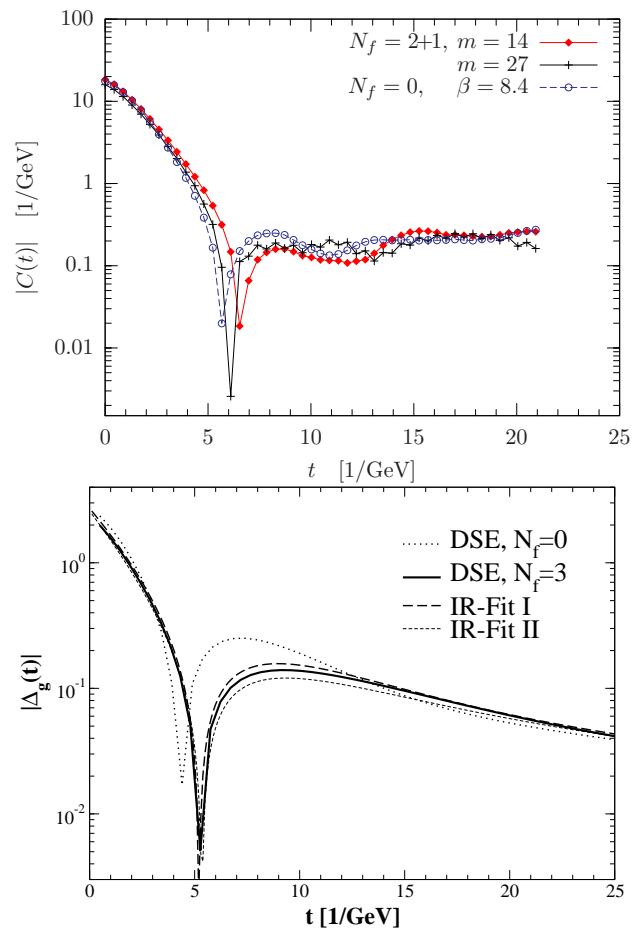}
\includegraphics[height=5.9cm]{dse_abs_ct}
\caption{The top figure corresponds to the absolute value of the gluon 
Schwinger function from our fine lattice calculations in quenched QCD as well 
as in full QCD. Error bars are not shown for simplicity. 
The bare quark masses for full QCD simulations are
${m=27}$~MeV and ${m=14}$ MeV, respectively.
The bottom figure is taken from~\cite{Alkofer:2003jj}. It shows the 
numerical results for the absolute value of the
Schwinger function from the DSE result compared to the fits in the infrared.
See Ref.~\cite{Alkofer:2003jj} for more details.}
\label{absschwinger}
\end{figure}

As mentioned in the introduction, some violations of positivity are a
regrettable consequence of the altered cut-off effects of improved
actions~\cite{Luscher:1984is}. These effects, being ultraviolet
effects, can cause violations of positivity in two-point functions at
short distances, particularly if simulations are performed too far
from the continuum limit.  Our study concerns the long distance
behavior of the gluon two-point function, and many preceding studies
(e.g.\ \cite{Bonnet:2001uh,Bonnet:2000kw}) have indicated that the
infrared behavior of the Landau-gauge gluon propagator is robust
against changes of the lattice spacing, at least for the range of
parameters we use. Therefore, our results are unlikely to be affected
by this problem. This claim is further supported by the calculation of
the Schwinger function on both coarse and fine lattices.  We find that
reflection positivity is clearly violated for both lattices; see, for example,
Fig.~\ref{fig:cmp_Ct_fine2coarse}.

At large Euclidean time, $C(t)$ is expected to reach zero from below. 
Unfortunately, this cannot be seen from our data shown in
Fig.~\ref{gp06}.
Something similar has been observed independently using a different 
fermion formulation~\cite{Sternbeck:2006cg}. The reason might be the  
difference in the number of lattice points in spatial and temporal directions. 
It is known~\cite{Leinweber:1998im,Leinweber:1998uu} that on an
asymmetric lattice the gluon propagator can violate the
infinite-volume continuum-limit tensor
structure (\eref{eq:Landau-Prop}) in the (extended) time direction  
at very low momentum values. One effect of this is to alter the
normalization of {$D(p=0)$} relative to $D(p>0)$. As is customary, we
assumed the continuum tensor structure and thus \mbox{$D(p=0)$} has
been normalized by $N_d=4$ and $D(p>0)$ by $N_d-1=3$. However, using a
slightly different normalization either for $D(0)$ or $D(p>0)$ the
data in Fig.~\ref{gp06} could be easily shifted towards larger values
such that (within errors) they equal zero at large~$t$. Therefore,
finite-volume effects could be responsible for $C(t)$ not reaching
zero from below at large~$t$.

Apart from $C(t)$, it is also interesting to consider its absolute
value and to compare our lattice data with corresponding
DSE results by Alkofer \emph{et al.}~\cite{Alkofer:2003jj}. Those
results have been reprinted on the right-hand side of
Fig.~\ref{absschwinger}.  On the left-hand side we present our lattice
result for $|C(t)|$ on the fine ensemble.  The two figures are remarkably 
alike: not only does the zero crossing occur at almost the same place in
Euclidean time, $t\approx 5~\text{GeV}^{-1}$ (about the size of a
hadron), but the DSE results also correctly predict the small shift
due to the inclusion of dynamical quarks.  This shift is largest in our data 
for the smallest light quark mass.  A similar result is obtained by using the
coarse lattices.

\section{Conclusions}

We have extended a previous lattice study \cite{Bowman:2004jm} of the
Landau-gauge gluon propagator in full QCD to a finer lattice. The
addition of quark loops has a clear, quantitative effect on the gluon
propagator, but its basic features are unaltered. Good scaling
behavior is observed for the gluon dressing function in both the quenched
and unquenched cases with these improved actions.

The violation of reflection positivity of the gluon propagator was
investigated by calculating the real space propagator, or Schwinger
function; the Landau gauge gluon propagator clearly violates
positivity in both quenched and full QCD.   

\section*{ACKNOWLEDGMENTS}

This research was supported by the Australian Research Council and by
grants of time on the Hydra Supercomputer, supported by the South
Australian Partnership for Advanced Computing.  We thank the MILC 
Collaboration for providing the results of their static quark potential 
measurements. J.\ B.\ Zhang is supported by Chinese national natural
science foundation grant 10675101.


\end{document}